\documentclass[conference]{IEEEtran}
\IEEEoverridecommandlockouts
\usepackage{cite}
\usepackage{amsmath,amssymb,amsfonts}
\usepackage{algorithmic}
\usepackage{graphicx}
\usepackage{textcomp}
\usepackage{xcolor}
\usepackage{hyperref}
\usepackage{braket}
\usepackage{amsthm}

\def\BibTeX{{\rm B\kern-.05em{\sc i\kern-.025em b}\kern-.08em
    T\kern-.1667em\lower.7ex\hbox{E}\kern-.125emX}}

\renewcommand{\figureautorefname}{Figure~\negthinspace}

\begin{document}

\title{Introduction to Quantum Machine Learning and Quantum Architecture Search \thanks{The views expressed in this article are those of the authors and do not represent the views of Wells Fargo. This article is for informational purposes only. Nothing contained in this article should be construed as investment advice. Wells Fargo makes no express or implied warranties and expressly disclaims all legal, tax, and accounting implications related to this article.}
}

\author{\IEEEauthorblockN{
Samuel Yen-Chi Chen\textsuperscript{1}  \ \
Zhiding Liang\textsuperscript{2}}
\IEEEauthorblockA{
\textsuperscript{1}Wells Fargo\\
\textsuperscript{2}Rensselaer Polytechnic Institute
\vspace{-0.15in}}
}

\maketitle

\begin{abstract}
Recent advancements in quantum computing (QC) and machine learning (ML) have fueled significant research efforts aimed at integrating these two transformative technologies. Quantum machine learning (QML), an emerging interdisciplinary field, leverages quantum principles to enhance the performance of ML algorithms. Concurrently, the exploration of systematic and automated approaches for designing high-performance quantum circuit architectures for QML tasks has gained prominence, as these methods empower researchers outside the quantum computing domain to effectively utilize quantum-enhanced tools. This tutorial will provide an in-depth overview of recent breakthroughs in both areas, highlighting their potential to expand the application landscape of QML across diverse fields.
\end{abstract}


\section{Introduction}
\begin{figure}[htbp]
\vskip -0.15in
\begin{center}
\includegraphics[width=1\columnwidth]{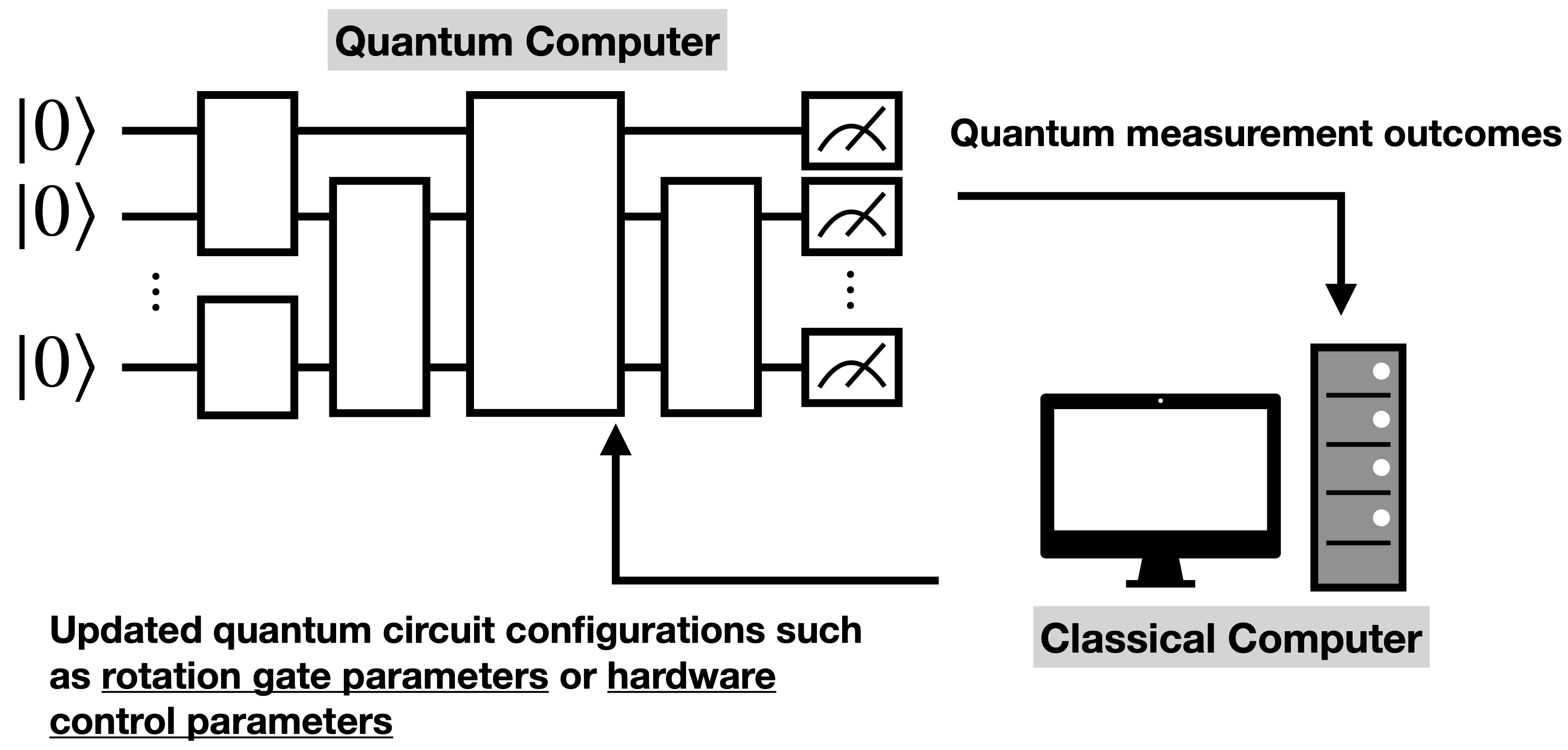}
\caption{{\bfseries Hybrid Quantum-Classical Computing.}}
\label{fig:hybrid_quantum_classical_scheme}
\end{center}
\vskip -0.15in
\end{figure}

Quantum computing (QC) offers the potential for substantial speedups in solving certain computationally challenging problems compared to classical computers. Recent advancements in quantum hardware, coupled with remarkable progress in classical AI and machine learning (ML) techniques, have sparked growing interest in merging these two technologies to further accelerate advancements in artificial intelligence.
Although current quantum computers are still limited by noise, a hybrid quantum-classical computing framework \cite{bharti2022noisy} has been proposed to harness the strengths of both quantum and classical computing resources. As depicted in \figureautorefname{\ref{fig:hybrid_quantum_classical_scheme}}, a computational task is divided into two components: tasks that benefit from quantum computational power are executed on quantum devices, while tasks where classical computing excels remain on classical systems.

\emph{Variational Quantum Algorithms} (VQAs) provide an efficient implementation of the hybrid quantum-classical framework for quantum machine learning (QML), where quantum circuit parameters are optimized using classical methods. These algorithms have demonstrated theoretical advantages \cite{abbas2021power,du2020expressive}. Empirically, hybrid approaches have been successfully applied to various domains, including classification \cite{chen2021end, qi2023qtnvqc, mitarai2018quantum, chen2022quantumCNN}, time-series prediction \cite{chen2022quantumLSTM}, natural language processing \cite{li2023pqlm, yang2022bert, di2022dawn, stein2023applying}, reinforcement learning \cite{chen2022variationalQRL,chen2020QRL,chen2024efficient,chen2023quantumLSTM_RL,skolik2021quantum,lockwood2020reinforcement,jerbi2021variational,CHEN2023321Async,coelho2024vqc}, and neural network model compression \cite{liu2024quantumTrain,liu2024qtrl,lin2024QT_LSTM,liu2024federated_QT,lin2024_QT_DeepFake}. 

In this tutorial, we will review these applications and discuss techniques for automatically discovering high-performance QML circuit architectures. This emerging field, known as \emph{quantum architecture search} (QAS), enables researchers outside the quantum computing domain to design effective QML models.
\section{Quantum Computing}
A \emph{qubit} serves as the fundamental building block of quantum information processing. In contrast to a classical \emph{bit}, which can exist only in the state $0$ or $1$, a qubit can simultaneously represent both $0$ and $1$ due to the phenomenon of quantum superposition. The quantum state of a single qubit is expressed as $ \ket{\Psi} = \alpha\ket{0} + \beta\ket{1} $, $\ket{0} = [1, 0]^{T}$ and $\ket{1} = [0, 1]^{T}$ are column vectors, and $\alpha$ and $\beta$ are complex coefficients from the set of complex numbers $\mathbb{C}$. For an $n$-qubit system, the corresponding state vector resides in a space of dimension $2^{n}$.
A \emph{quantum gate}, represented by $U$, performs a transformation on a quantum state $\ket{\Psi}$, resulting in a new state $\ket{\Psi'}$ according to the operation $ \ket{\Psi'} = U \ket{\Psi} $. These gates correspond to \emph{unitary transformations}, which satisfy the condition $ U U^{\dagger} = U^{\dagger} U = \mathbb{I}_{2^{n} \times 2^{n}} $, where $n$ denotes the number of qubits.
It has been shown that we only need a small set of fundamental quantum gates to achieve universal quantum computation. One such set includes single-qubit gates $H$, $\sigma_x$, $\sigma_y$, $\sigma_z$, $R_{x}(\theta) = e^{-i\theta\sigma_x/2}$, $R_{y}(\theta) = e^{-i\theta\sigma_y/2}$, $R_{z}(\theta) = e^{-i\theta\sigma_z/2}$, and the two-qubit gate CNOT. 
In quantum machine learning (QML), the rotation gates $R_x$, $R_y$, and $R_z$ play a crucial role, as their rotation angles can be used as tunable parameters during optimization. For quantum operations involving multi-qubit systems, the overall unitary transformation is formed by taking the tensor product of individual single-qubit or two-qubit operations, written as $ U = U_{1} \otimes U_{2} \otimes \cdots \otimes U_{k} $, where each $ U_{j} $ represents a specific single-qubit or two-qubit gate. A sequence of such unitary transformations can be applied consecutively to evolve a quantum state $ \ket{\Psi} $, which corresponds to performing matrix multiplication.
At the final stage of executing a quantum circuit, a process known as \emph{measurement} is performed to extract information from the quantum system for further analysis or processing. Each execution of the quantum circuit yields a binary string. By repeating this process multiple times, it becomes possible to estimate the probabilities of different computational basis states (e.g., $\ket{0,\cdots,0}$, $\cdots$, $\ket{1,\cdots,1}$) or to compute expectation values corresponding to specified observables (e.g., Pauli $X$, $Y$, and $Z$ operators).
\section{Quantum Neural Networks}
\begin{figure}[htbp]
\vskip -0.15in
\begin{center}
\includegraphics[width=1\columnwidth]{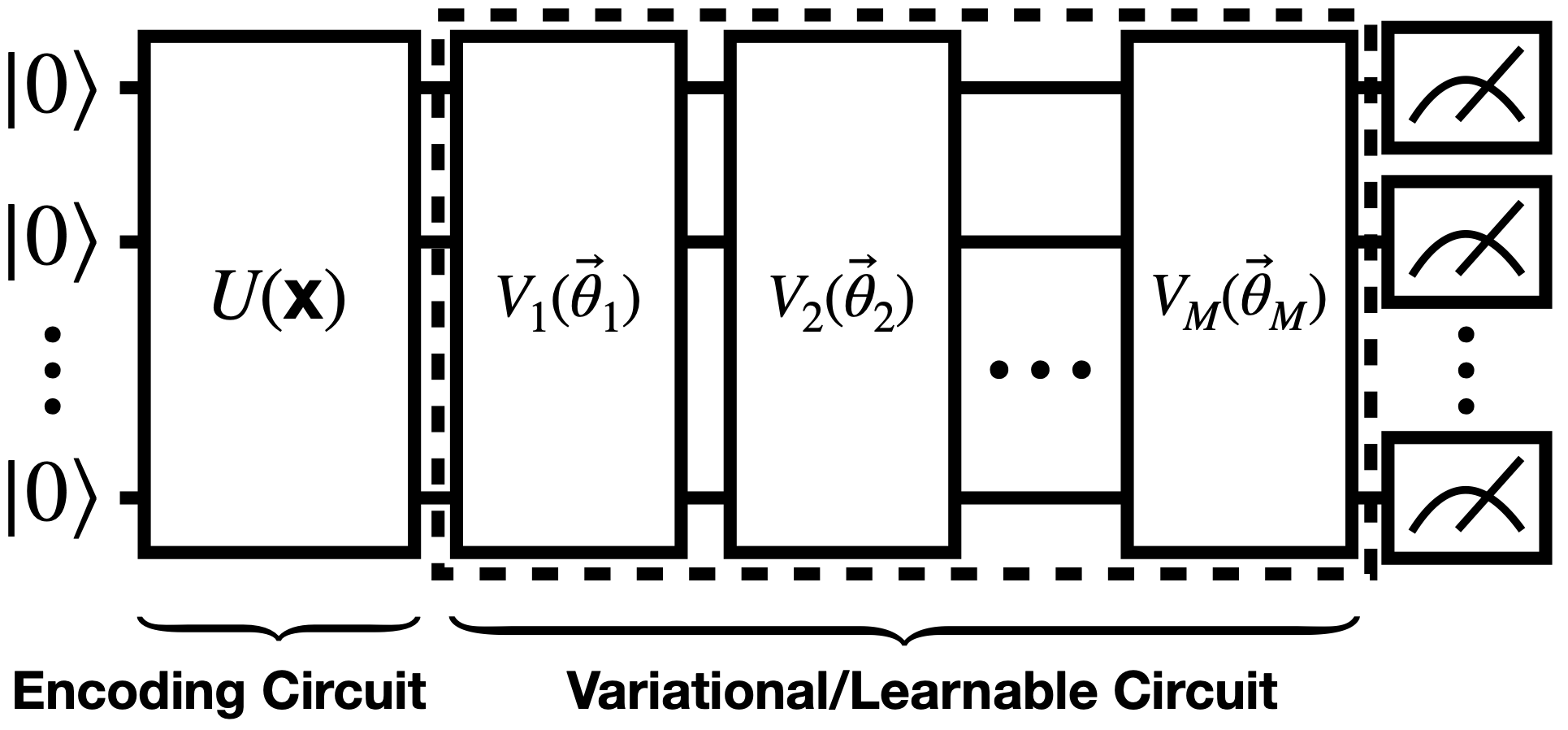}
\caption{{\bfseries Generic architecture of a variational quantum circuit (VQC).}}
\label{fig:generic_vqc}
\end{center}
\vskip -0.15in
\end{figure}

A Variational Quantum Circuit (VQC), also known as a Parameterized Quantum Circuit (PQC), and commonly used as a Quantum Neural Network (QNN), typically consists of three main components: the \emph{encoding circuit}, the \emph{parameterized or variational circuit}, and the final \emph{quantum measurement}.
The encoding circuit, denoted as $U(\vec{x})$, maps the input vector of classical numerical values $\vec{x}$ into a quantum state by transforming it into $U(\vec{x})\ket{0}^{\otimes n}$, where $\ket{0}^{\otimes n}$ represents the ground state of the quantum system, and $n$ is the number of qubits.
The parameterized circuit processes and transforms the encoded quantum state, resulting in $W(\Theta)U(\vec{x})\ket{0}^{\otimes n}$. Typically, the variational (parameterized or learnable) circuit $W(\Theta)$ is composed of multiple layers of trainable sub-circuits $V_{j}(\vec{\theta_{j}})$, as illustrated in \figureautorefname{\ref{fig:generic_vqc}}, and is expressed as $W(\Theta) = \prod_{j = M}^{1} V_{j}(\vec{\theta_{j}})$, where $\Theta$ denotes the set of all learnable parameters $\{\vec{\theta_{1}}, \cdots, \vec{\theta_{M}}\}$.
Then the quantum state generated by the encoding and parameterized circuit can be shown as, 
\begin{equation}
\label{eqn:vqc_state_psi}
    \ket{\Psi} = W(\Theta) U(\vec{x})\ket{0}^{\otimes n} = \left( \prod_{j = M}^{1} V_{j}(\vec{\theta_{j}}) \right) U(\vec{x})\ket{0}^{\otimes n}
\end{equation}
Information from the QNN or VQC is obtained by performing measurements on the quantum state using predefined observables, denoted as $\hat{B}_{k}$.
The full operation of a QNN or VQC can be seen as a quantum function $\overrightarrow{f(\vec{x} ; \vec{\theta})}=\left(\left\langle\hat{B}_1\right\rangle, \cdots,\left\langle\hat{B}_n\right\rangle\right)$, where $\left\langle\hat{B}_{k}\right\rangle =\left\langle 0\left|U^{\dagger}(\vec{x})W^{\dagger}(\Theta) \hat{B}_{k} W(\Theta)U(\vec{x})\right| 0\right\rangle$. 
Expectation values $\left\langle\hat{B}_{k}\right\rangle$ can be obtained by performing repeated measurements (shots) on quantum hardware or by direct computation using quantum simulation software on classical computers. The observable $\hat{B}_{k}$ is usually a predefined Hermitian matrix, with the Pauli-$Z$ matrix being a commonly used choice.
\section{Quantum Classification}
QML models are commonly applied to traditional classification tasks. The seminal work \cite{mitarai2018quantum} demonstrated the feasibility of training quantum circuit parameters for classifying simple datasets using the gradient descent approach. Subsequent advancements improved upon this framework by stacking multiple layers of trainable quantum circuits, enabling more complex classification tasks with models such as quantum convolutional neural networks (QCNN), as described in \cite{chen2022quantumCNN}. To address input data with feature dimensions that exceed the capacity of current quantum devices or simulators, hybrid models that incorporate learnable classical feature extractors have also been developed \cite{chen2021end,qi2023qtnvqc}.
\section{Federated Quantum Machine Learning}
\begin{figure}[htbp]
\vskip -0.15in
\begin{center}
\includegraphics[width=1\columnwidth]{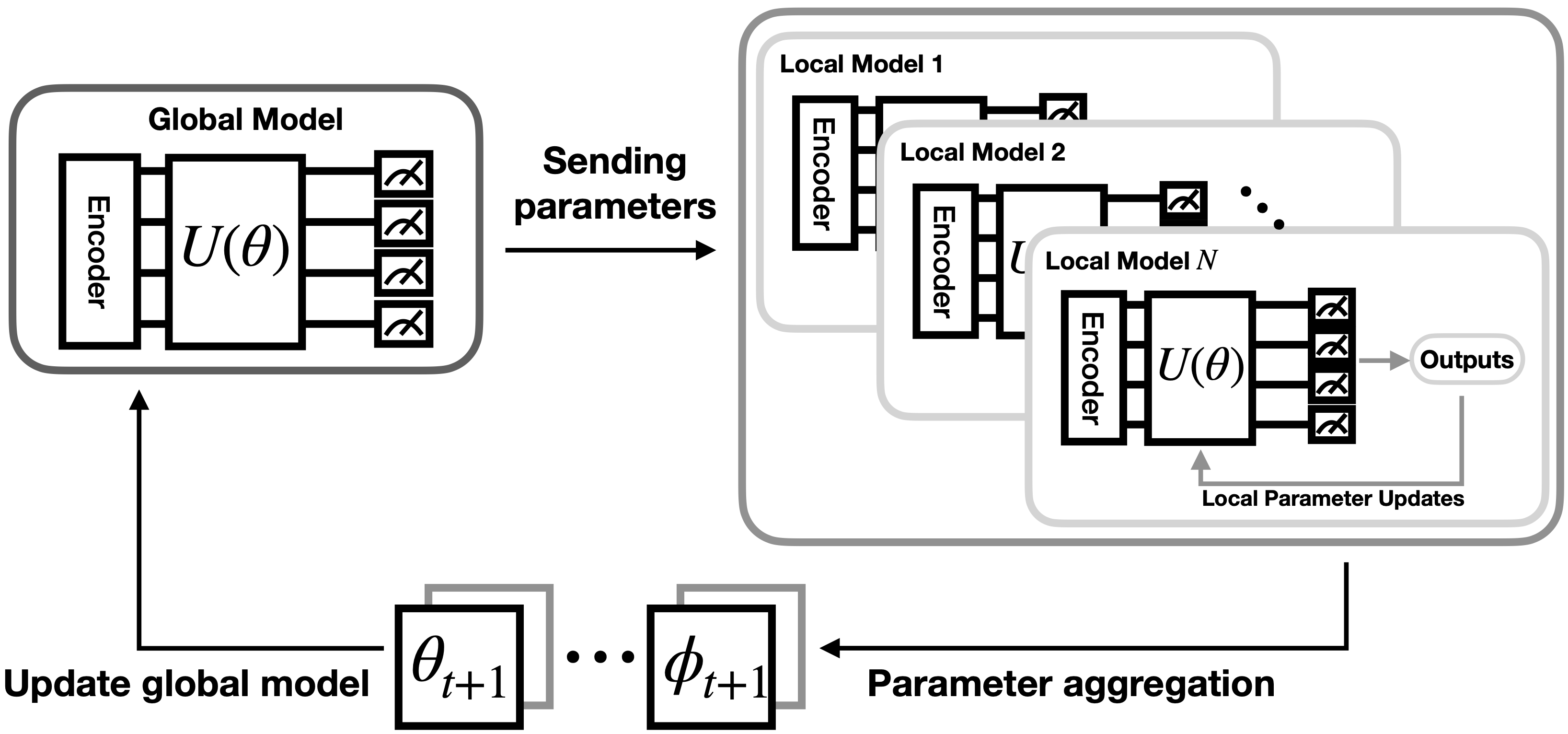}
\caption{{\bfseries Federated Quantum Machine Learning.}}
\label{fig:fed_qml}
\end{center}
\vskip -0.15in
\end{figure}

Federated learning (FL) has gained prominence due to increasing privacy concerns surrounding the use of large-scale datasets and cloud-based deep learning. The key components of a federated learning process include a \emph{central node} and multiple \emph{client nodes}. The central node maintains the \emph{global model} and receives trained parameters from the client devices. It then performs an \emph{aggregation} process to update the global model and distributes the updated model to all client nodes. Client nodes perform local training using the received model and their own private datasets, which are typically subsets of the overall data. In the QML setting \cite{chen2021federated,chehimi2022quantum}, one can envision a network of smaller quantum computers, each holding private or sensitive data. These local participants train their QML models using their own datasets, where the models may be hybrid—comprising both classical and quantum components. After completing local training, the participants upload their model parameters to the central server, where an aggregated global model is created and shared across the network. The concept of federated QML is illustrated in \figureautorefname{\ref{fig:fed_qml}}.
\section{Quantum Recurrent Neural Networks}
\begin{figure}[htbp]
\vskip -0.15in
\begin{center}
\includegraphics[width=1\columnwidth]{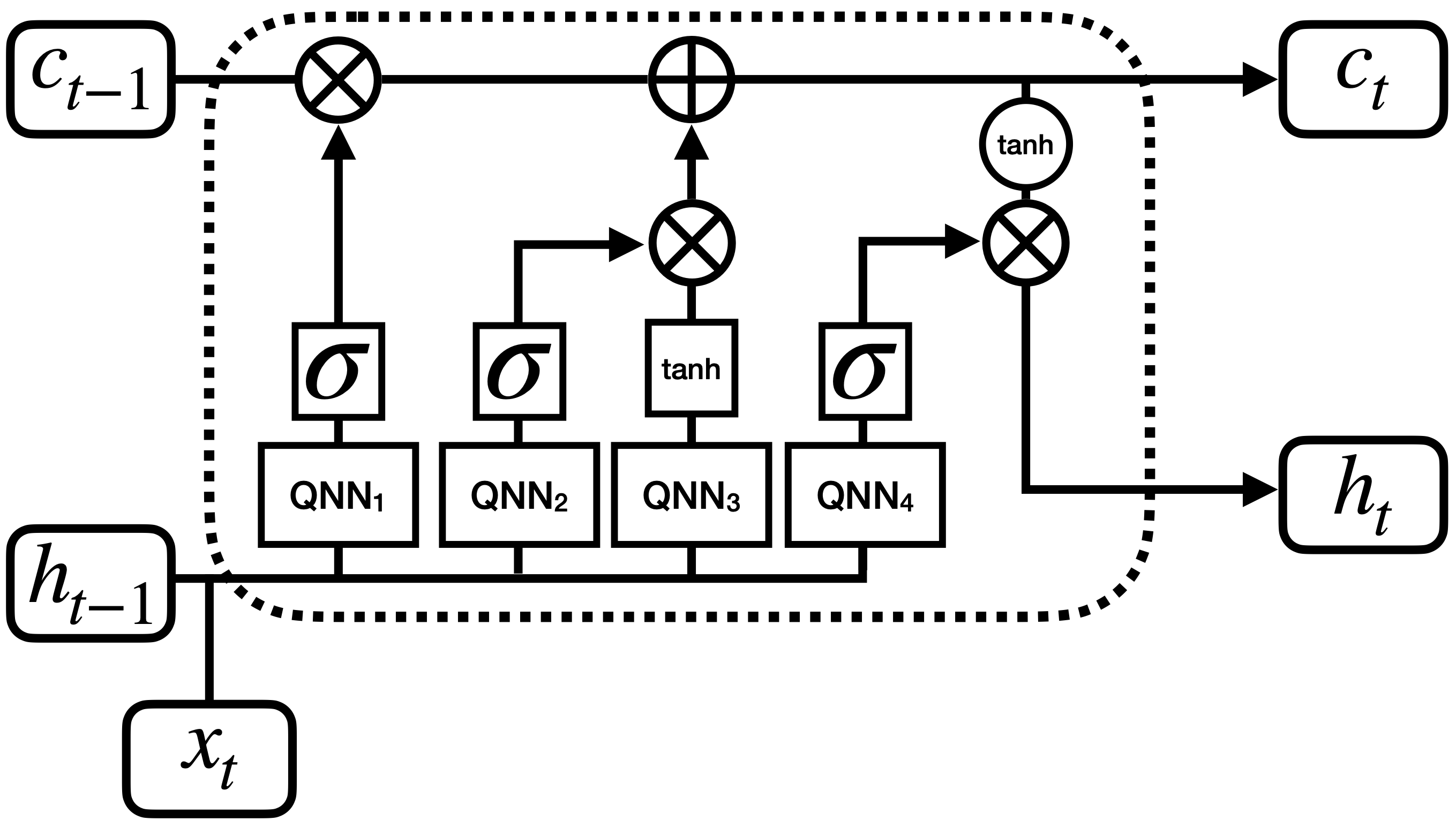}
\caption{{\bfseries Quantum Long Short-term Memory (QLSTM).}}
\label{fig:qlstm}
\end{center}
\vskip -0.15in
\end{figure}

The vanilla QNN functions similarly to a feed-forward neural network. However, some ML tasks require the model to retain information from previous time steps. In classical ML, recurrent neural networks (RNNs), including long short-term memory (LSTM) networks \cite{hochreiter1997long}, were introduced to handle tasks involving temporal dependencies, such as time-series prediction and natural language processing. The quantum equivalent, referred to as the quantum LSTM (QLSTM), extends this concept by replacing classical neural networks with QNNs \cite{chen2022quantumLSTM}, as depicted in \figureautorefname{\ref{fig:qlstm}}.
A formal mathematical formulation of a QLSTM cell is given by,
\begin{subequations}
\allowdisplaybreaks
    \begin{align}
    f_{t} &= \sigma\left(\text{QNN}_{1}(v_t)\right) \label{eqn:qlstm-f}\\
    i_{t} &= \sigma\left(\text{QNN}_{2}(v_t)\right) \label{eqn:qlstm-i}\\ 
    \tilde{C}_{t} &= \tanh \left(\text{QNN}_{3}(v_t)\right) \label{eqn:qlstm-bigC}\\
    c_{t} &= f_{t} * c_{t-1} + i_{t} * \tilde{C}_{t} \label{eqn:qlstm-c}\\
    o_{t} &= \sigma\left(\text{QNN}_{4}(v_t)\right) \label{eqn:qlstm-o}\\ 
    h_{t} &= o_{t} * \tanh \left(c_{t}\right)\label{eqn:qlstm-h}
    \end{align}
    \label{eqn:qlstm}
\end{subequations}
where $v_t=\left[h_{t-1} x_{t}\right]$ represents the concatenation of input $x_{t}$ at time-step $t$ and the hidden state $h_{t-1}$ from the previous time-step $t-1$. 
Studies have shown that QLSTM can outperform classical LSTM in various time-series prediction tasks, especially when the model sizes (number of trainable parameters) are similar \cite{chen2022quantumLSTM}. Beyond time-series prediction, QLSTM has also demonstrated effectiveness in natural language processing \cite{li2023pqlm,di2022dawn,stein2023applying} and reinforcement learning within partially observable environments \cite{chen2023quantumLSTM_RL}.

\begin{figure}[htbp]
\vskip -0.15in
\begin{center}
\includegraphics[width=1\columnwidth]{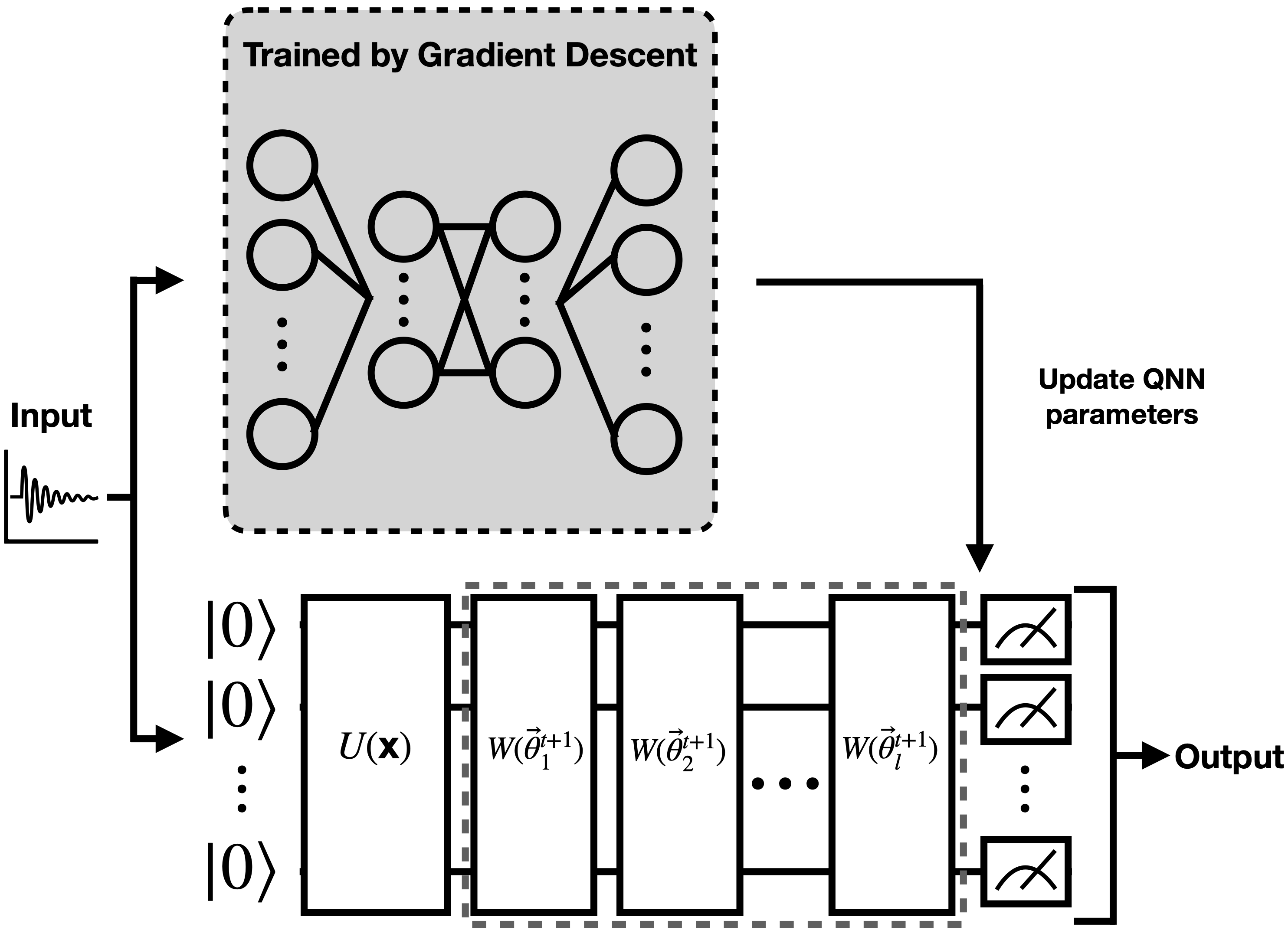}
\caption{{\bfseries Quantum Fast Weights Programmer (QFWP).}}
\label{fig:qfwp}
\end{center}
\vskip -0.15in
\end{figure}

Although RNN-based models are effective for handling sequential data and dynamic environments, a significant challenge in scaling quantum RNNs lies in the need to compute gradients through deep computational graphs, a process known as backpropagation-through-time (BPTT). An alternative approach to harness the potential of quantum RNNs without the computational overhead of BPTT is to treat RNNs as \emph{reservoirs}. These quantum RNN reservoirs are randomly initialized and remain untrained. While this approach significantly reduces the required training resources, numerical studies have shown that QLSTM reservoirs can achieve performance comparable to fully trained QLSTMs in certain reinforcement learning tasks \cite{chen2024efficient}.
An alternative approach for building QML models capable of capturing temporal or sequential dependencies without relying on recurrent connections in quantum neural networks is the \emph{Quantum Fast Weights Programmer} (QFWP). The original idea of \emph{Fast Weight Programmers} (FWP) was introduced by Schmidhuber in his early work \cite{schmidhuber1992learning}.
In this sequential learning framework, two separate neural networks are utilized: the \emph{slow programmer} and the \emph{fast programmer}. Within this approach, the neural network weights act as the \emph{program} for the model or agent. The core concept of the FWP is that the slow programmer generates \emph{updates} or \emph{modifications} to the fast programmer's network weights based on the observations at each time step.
This \emph{reprogramming} process quickly directs the fast programmer's attention to relevant information in the incoming data stream. Notably, the slow programmer does not completely overwrite the fast programmer's model weights but applies targeted updates instead (e.g. add some values to the original weights). This approach allows the fast programmer to retain information from past observations, enabling a simple feed-forward neural network to manage sequential prediction or complex control tasks without the substantial computational overhead associated with RNNs.
The concept of FWP can be extended to the hybrid quantum-classical domain, as described in \cite{chen2024learning}. In this quantum adaptation, classical neural networks are used to construct the \emph{slow} networks, which generate updates for the parameters of the \emph{fast} networks implemented as a VQC, as illustrated in \figureautorefname{\ref{fig:qfwp}}. 
In this framework, only the weights of the slow programmer are updated using conventional gradient descent optimization.
\section{Quantum Reinforcement Learning}
\begin{figure}[htbp]
\vskip -0.15in
\begin{center}
\includegraphics[width=1\columnwidth]{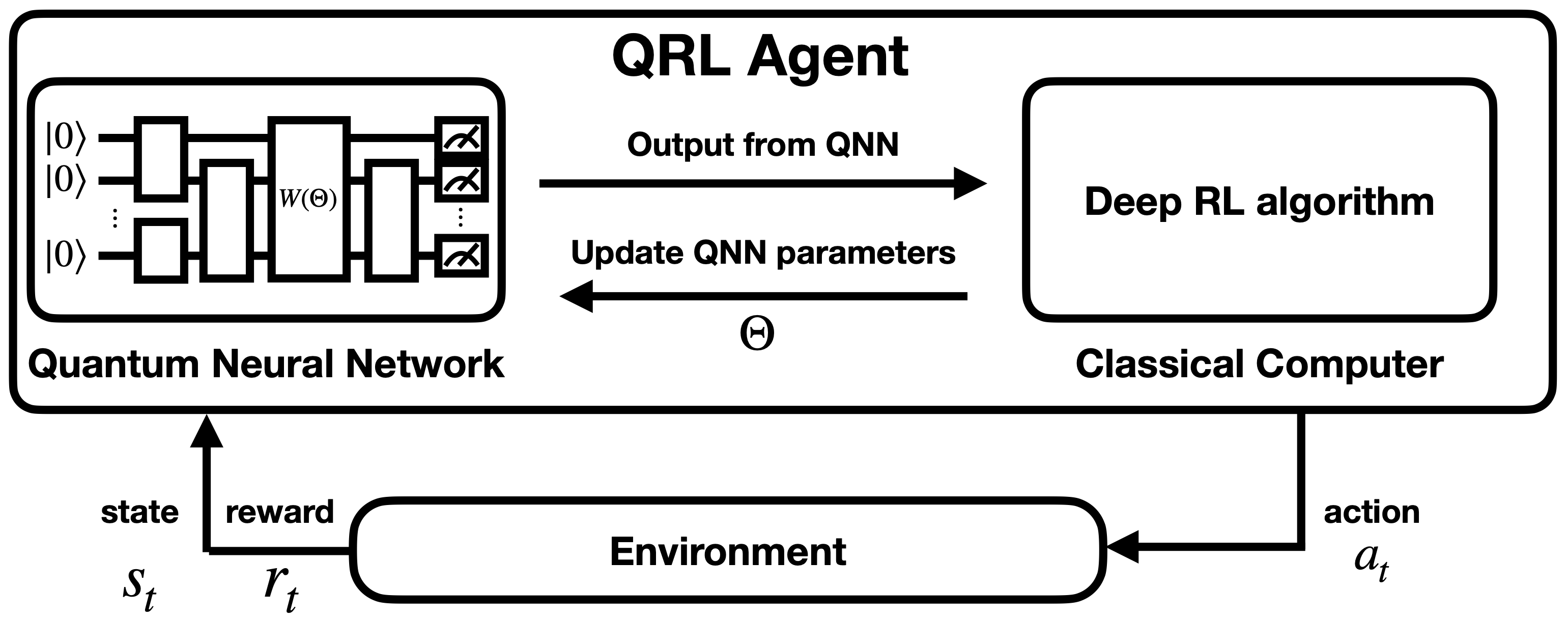}
\caption{{\bfseries Concept of quantum reinforcement learning (QRL).}}
\label{fig:qrl}
\end{center}
\vskip -0.2in
\end{figure}

\emph{Reinforcement Learning} (RL) is a fundamental paradigm in machine learning, wherein an autonomous entity, known as the \emph{agent}, learns to make decisions by interacting with its environment through an iterative process.
Although the concept is straightforward, RL has proven capable of solving complex sequential decision-making tasks, such as achieving human-level or superhuman performance in playing computer games.
The agent interacts with a structured \emph{environment} $\mathcal{E}$ over discrete time steps. At each time step $t$, the environment provides the agent with a \emph{state} or \emph{observation} $s_t$. Based on this information, the agent selects an \emph{action} $a_t$ from the available action set $\mathcal{A}$ according to a \emph{policy} $\pi$. The policy $\pi$ defines how the agent maps the observed state $s_t$ to an action $a_t$. If the policy is stochastic, the action $a_t$ is sampled from a probability distribution $\pi(a_t|s_t)$ rather than being chosen deterministically.
Once the agent executes the action $a_t$, it transitions to a new state $s_{t+1}$ and receives a scalar \emph{reward} $r_t$. This cycle repeats until the agent either reaches a terminal state or satisfies a predefined stopping condition, such as reaching a maximum number of steps. The sequence of interactions, beginning from the initial state and continuing until termination, is known as an \emph{episode}. \emph{Quantum reinforcement learning} (QRL) incorporates quantum neural networks to learn policy or value functions within this framework. The idea of QRL is illustrated in \figureautorefname{\ref{fig:qrl}}.
The first VQC-based QRL approach was introduced in \cite{chen2020QRL}, focusing on environments with discrete observation spaces, such as Frozen Lake and Cognitive Radio. This type of QRL is inspired by the original deep $Q$-learning framework. Subsequent advancements in quantum deep $Q$-learning have addressed environments with continuous observation spaces \cite{lockwood2020reinforcement, skolik2021quantum}, demonstrating notable advantages. Beyond value-based QRL, policy-based QRL has also seen significant progress, including approaches such as REINFORCE \cite{jerbi2021variational}, Advantage Actor-Critic (A2C) \cite{kolle2024quantum}, and Asynchronous Advantage Actor-Critic (A3C) \cite{CHEN2023321Async}. To handle the complexities of partially observable environments, researchers have explored quantum recurrent neural networks, such as QLSTM, as RL policies \cite{chen2023quantumLSTM_RL,chen2024efficient}. Additionally, the QFWP model, described in the previous section, has been employed to solve reinforcement learning tasks that require handling long-term dependencies \cite{chen2024learning}.
\section{Quantum Neural Networks as Model Compressor}
\begin{figure}[htbp]
\vskip -0.15in
\begin{center}
\includegraphics[width=1\columnwidth]{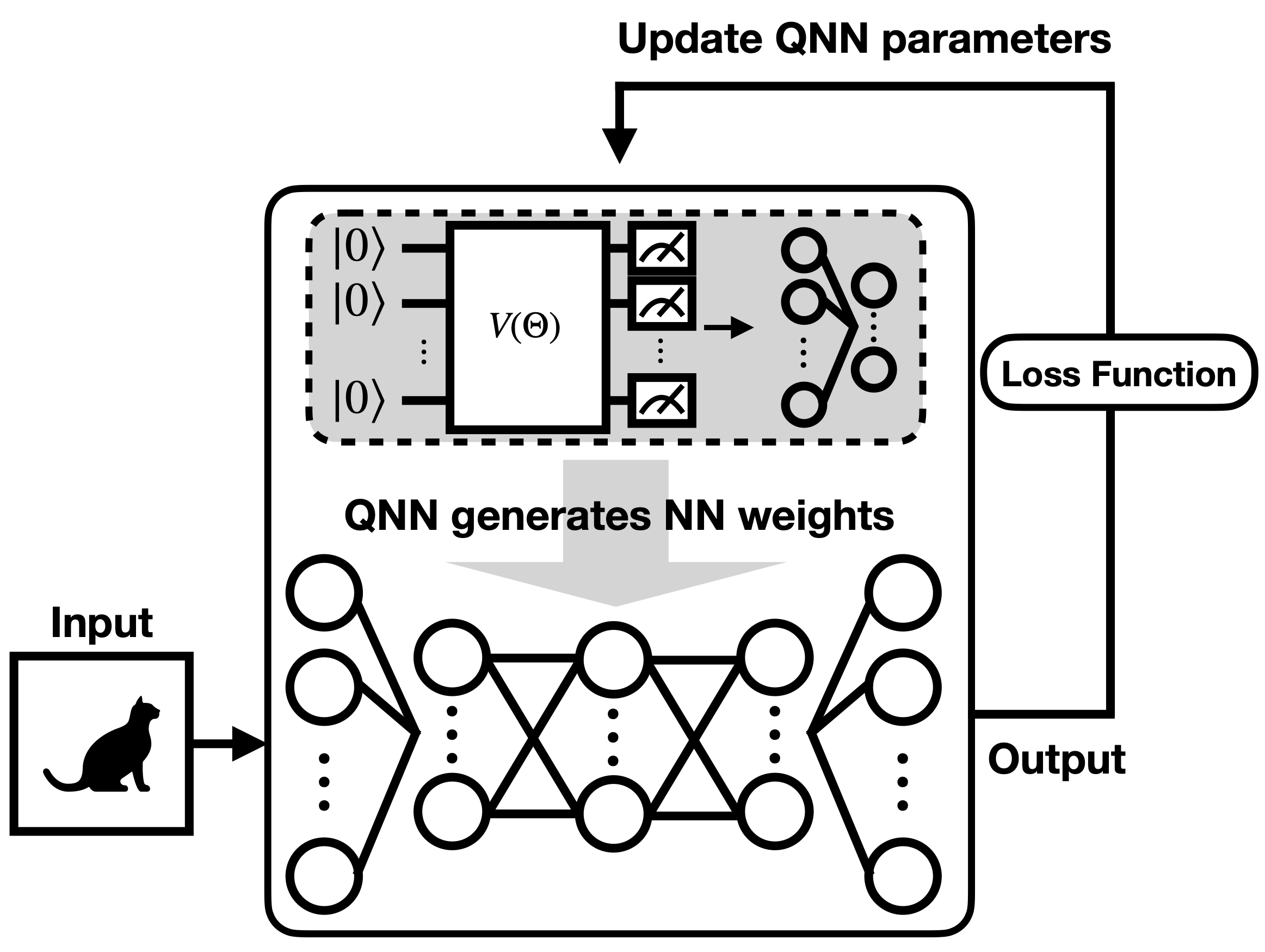}
\caption{{\bfseries Concept of Quantum-Train (QT).}}
\label{fig:qt}
\end{center}
\vskip -0.15in
\end{figure}

A key challenge for QML applications is the limited availability of quantum resources, which increases the complexity of both training and inference processes. In \cite{liu2024quantumTrain}, the authors propose a quantum-enhanced model compression or generation method called Quantum-Train (QT), where a QNN is utilized to generate the weights for a classical neural network (NN), as shown in \figureautorefname{\ref{fig:qt}}. For an $N$-qubit QNN, measuring the expectation values of individual qubits provides up to $N$ values. However, by collecting the probabilities of all computational basis states $\{\ket{00 \cdots 0}, \cdots, \ket{11 \cdots 1}\}$, a total of $2^{N}$ values can be obtained. These values are rescaled and assigned as NN weights. Therefore, for an NN with $M$ weights, only $\lceil \log_{2}M \rceil$ qubits are required to generate the weights. Numerical experiments demonstrate that the quantum circuit can efficiently generate NN weights, achieving inference performance comparable to conventional training methods. This quantum-enhanced model compression has been successfully applied in various domains, including classification \cite{liu2024federated_QT,lin2024_QT_DeepFake}, sequence learning \cite{lin2024QT_LSTM}, and reinforcement learning \cite{liu2024qtrl}.
\section{Quantum Architecture Search}
\begin{figure}[htbp]
\vskip -0.15in
\begin{center}
\includegraphics[width=1\columnwidth]{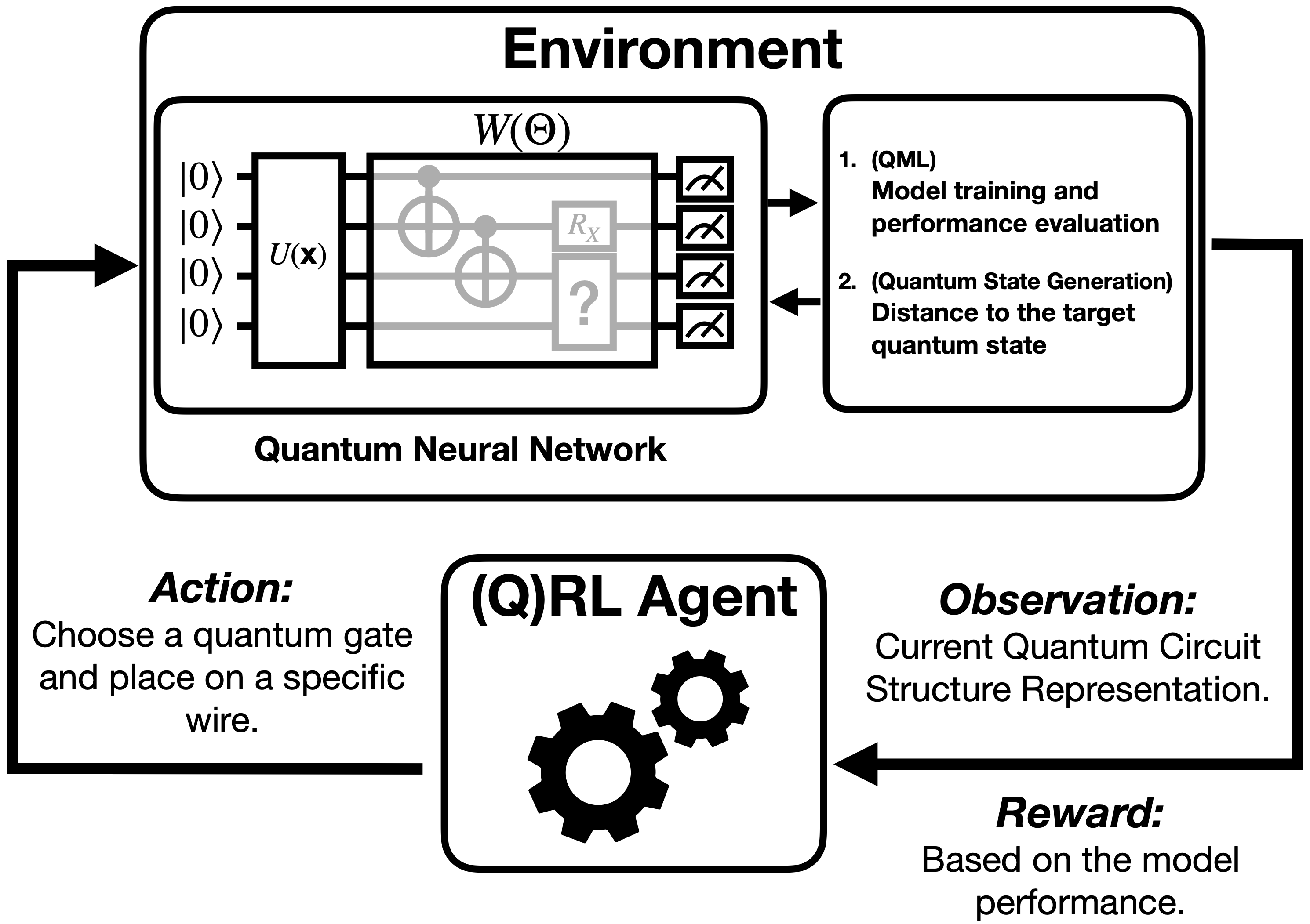}
\caption{{\bfseries Quantum Architecture Search (QAS) as a RL problem.}}
\label{fig:QAS_RL}
\end{center}
\vskip -0.15in
\end{figure}

\begin{figure}[htbp]
\vskip -0.15in
\begin{center}
\includegraphics[width=1\columnwidth]{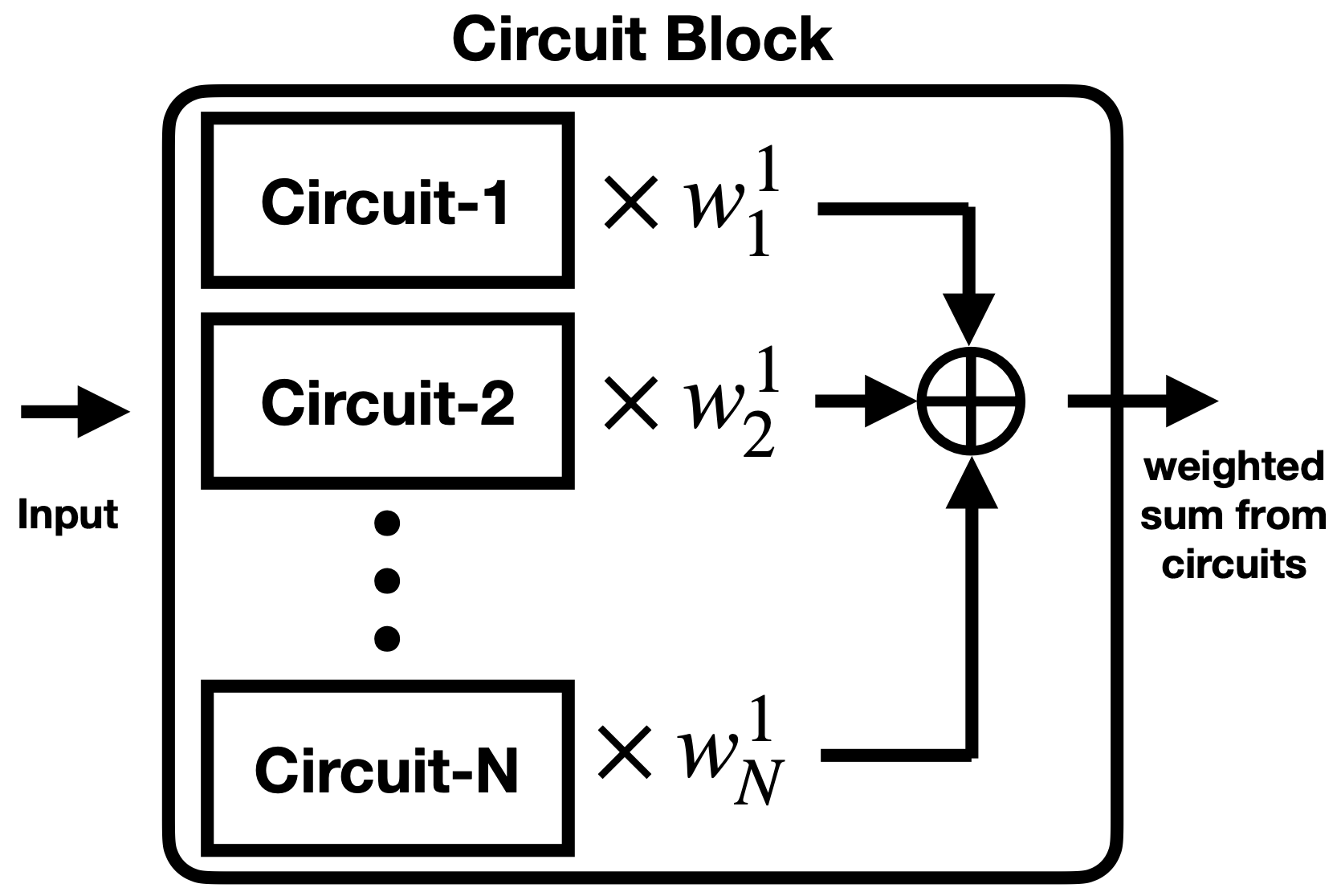}
\caption{{\bfseries Differentiable Quantum Architecture Search (DiffQAS) with learnable structural weights.}}
\label{fig:DiffQAS_Blocks}
\end{center}
\vskip -0.15in
\end{figure}

Designing an effective QML model for a specific task requires expertise in quantum information science. To broaden the scope of QML applications, considerable efforts have been made to develop automated procedures for constructing high-performance quantum circuit models. These methods fall under the field of Quantum Architecture Search (QAS) \cite{martyniuk2024quantum}, which utilizes various search algorithms and machine learning techniques to generate quantum architectures for designated tasks.
For any QML task—be it classification, time-series prediction, or reinforcement learning (RL)—there are specific metrics used to evaluate model performance. Common metrics include classification accuracy, prediction errors for time-series tasks, and total returns in RL scenarios. These performance indicators can be combined with other factors, such as circuit depth or the number of specific quantum gates, to assess the quality of a discovered QML circuit architecture. By iteratively performing these evaluations, an effective QAS algorithm can identify a high-performance QML circuit tailored to the specific task.
Various approaches have been explored to address the challenges in QAS, including RL \cite{kuo2021quantum,ye2021quantum,zhu2023quantum,sogabe2022model,wang2024rnn,fodera2024reinforcement,chen2023quantumRL_QAS,dai2024quantum}, evolutionary algorithms \cite{ding2022evolutionary,chen2024evolutionary}, and differentiable programming \cite{zhang2022differentiable,sun2023differentiable,chen2024differentiable}.
One approach to performing QAS is through evolutionary search. This method involves defining a specific type of \emph{quantum circuit representation}, referred to as a \emph{chromosome}, which encodes the quantum circuit architecture in a standard data structure. The representation can take the form of a string, array, dictionary, or other data structure, as long as it can be easily modified (i.e., \emph{mutated}) and mapped back to a valid quantum circuit for evaluation. For instance, in \cite{ding2022evolutionary}, evolutionary algorithms are used to discover high-performance circuits for QRL. The authors define a set of candidate VQC blocks, including entangling blocks, data-encoding blocks, variational blocks, and measurement blocks. The goal of the evolutionary search is to identify the optimal sequence of these blocks while adhering to a constraint on the maximum number of circuit blocks. Evolutionary search methods can be applied not only to identify QML models for predefined tasks such as QRL but also to discover QNN architectures with specific properties, such as \emph{model capacity} or \emph{effective dimension}, as described in \cite{chen2024evolutionary}.
Since QAS can be modeled as a sequential decision-making process, RL provides a natural solution. As illustrated in \figureautorefname{\ref{fig:QAS_RL}}, the RL-based QAS framework \cite{dai2024quantum,kuo2021quantum,ye2021quantum,chen2023quantumRL_QAS} operates as follows: given a target task—such as a QML application or quantum state generation—and a set of hardware constraints (e.g., the maximum number of quantum operations, permissible quantum gates, or hardware connectivity), the goal is to train an RL agent to sequentially select quantum operations and place them on appropriate qubits or circuit wires. After each action, the environment—either a physical quantum device or a quantum simulator—evaluates the current circuit using a predefined performance metric, such as model accuracy or quantum state fidelity, and returns a reward to the agent. The agent uses this reward, along with observations from the environment (e.g., measured observables from the constructed quantum circuit), to refine its parameters through standard RL optimization methods.
A key challenge in both RL and evolutionary algorithms is the need to evaluate circuit performance each time a specific architecture is selected. This evaluation becomes increasingly complex as the search space grows, such as when additional qubits or more allowed circuit components are introduced. Moreover, these methods require tuning multiple hyperparameters, such as mutation rates and crossover probabilities for evolutionary algorithms, and exploration-exploitation balances for RL techniques. Differentiable search has emerged as an alternative approach to address these challenges. Differentiable programming has been applied in QAS for various QML tasks, including optimization, classification, and reinforcement learning \cite{zhang2022differentiable,sun2023differentiable,chen2024differentiable}. One major advantage of differentiable QAS is that the parameters governing the quantum circuit architecture are optimized simultaneously with the parameters of the QNNs/VQCs (e.g., rotation angles). Additionally, differentiable search requires fewer hyperparameters than evolutionary or RL-based methods, thereby simplifying the optimization process.
As illustrated in \figureautorefname{\ref{fig:DiffQAS_Blocks}}, consider the task of constructing a quantum circuit $\mathcal{C}$, which consists of multiple sub-components $\mathcal{S}_{1}, \mathcal{S}_{2}, \dots, \mathcal{S}_{n}$. Each sub-component $\mathcal{S}_{i}$ is associated with a set of available circuit options $\mathcal{B}_{i}$, where $|\mathcal{B}_{i}|$ denotes the number of feasible choices for that sub-component. Consequently, the total number of possible configurations for the circuit $\mathcal{C}$ is given by $N = |\mathcal{B}_{1}| \times |\mathcal{B}_{2}| \times \dots \times |\mathcal{B}_{n}|$. A naive approach to this problem would involve enumerating all possible circuit configurations and evaluating their performance individually. However, this brute-force method becomes impractical as the number of possible circuits grows. Instead, all possible circuits can be parameterized, and their realizations can be evaluated in parallel to compute the ensemble output as $f_{\mathcal{C}} = \sum_{j = 1}^{N} w_{j} f_{\mathcal{C}_{j}}$, where $w_{j}$ represents the trainable structural weight for the $j$-th circuit realization, and $f_{\mathcal{C}_{j}}$ denotes the output of the $j$-th circuit. The structural weights $w_{j}$ are optimized concurrently with the conventional QNN parameters, such as rotation angles, using gradient-based methods.
\section{Conclusion and Outlook}
In this tutorial, we present an overview of various quantum machine learning (QML) applications, including classification, sequential learning, natural language processing, and reinforcement learning. Additionally, we explore the use of quantum neural networks (QNNs) as model compressors for classical neural networks by leveraging the exponentially large quantum Hilbert space. We also highlight several quantum architecture search approaches, such as evolutionary optimization, reinforcement learning-driven methods, and differentiable programming techniques. The integration of quantum and classical methods holds significant promise for impactful advancements in the coming years.
\clearpage
\bibliographystyle{IEEEtran}
\bibliography{bib/qml_examples,bib/qas,bib/qt,bib/qc,bib/vqc,bib/classical_ml}

\end{document}